\documentclass[twocolumn,preprintnumbers,amsmath,amssymb]{revtex4}

\usepackage{graphicx}
\usepackage{dcolumn}
\usepackage{bm}

\begin{document}

\title{Does Bohm's Quantum Force Have a Classical Origin?
}


\author{David C. Lush}


\affiliation{%
 d.lush@comcast.net \\}%


\date{\today}

\begin{abstract}

In the de Broglie - Bohm formulation of quantum mechanics, the electron is stationary in the ground state of the hydrogen atom, because the quantum force exactly cancels the Coulomb attraction of the electron to the proton.  In this paper it is shown that classical electrodynamics similarly predicts the Coulomb force can be effectively canceled by part of the magnetic force that occurs between two similar particles each consisting of a point charge moving with circulatory motion at the speed of light.  Supposition of such motion is the basis of the {\em Zitterbewegung} interpretation of quantum mechanics.  The magnetic force between two luminally-circulating charges for separation large compared to their circulatory motions contains a radial inverse square law part with magnitude equal to the Coulomb force, sinusoidally modulated by the phase difference between the circulatory motions. When the particles have equal mass and their circulatory motions are aligned but out of phase, part of the magnetic force is equal but opposite the Coulomb force. This raises a possibility that the quantum force of Bohmian mechanics may be attributable to the magnetic force of classical electrodynamics. It is further shown that non-relativistic relative motion between the particles leads to modulation of the magnetic force with spatial period equal to the de Broglie wavelength.

\end{abstract}

\maketitle

\section{Introduction}
\label{intro}

The quantum potential of the de Broglie - Bohm formulation of quantum mechanics \cite{Bohm1952a} implies existence of a quantum force, which can be viewed as embodying the difference between classical and quantum mechanics. Where a na\"ive  application of classical mechanics to the hydrogen atom obtains Kepler-like orbital motion with radiative decay, in the de Broglie - Bohm theory the quantum force in the ground state exactly cancels the Coulomb attraction of the electron towards the proton.  Undergoing no acceleration, the electron will not radiate, and so the ground state does not decay. 

A plausible analogue of the quantum force of Bohmian mechanics is not obvious in classical electrodynamics.  If electric forces are accounted for by the Coulomb interaction, then only magnetic forces remain, which are generally regarded as weak in atomic physics compared to the Coulomb force. Nonetheless, the purpose of the present paper is to propose that magnetic forces expected if the {\em zitterbewegung} interpretation of quantum mechanics \cite{Hestenes1990} is correct may provide a basis for explaining the quantum force as a consequence of classical electrodynamics.

\section{Classical Spinning Particle Model}
\label{sectzbwinterpqm}

Since publication of the Dirac theory of the electron \cite{Dirac1928}, it has been clear that a classical current loop is not a good representation of the electron.  Although in the semiclassical analysis of the spin-orbit coupling the electron is modeled as an ideal magnetic dipole as associated with a classsical current loop,  in the Dirac theory, unlike a current loop, the electron has a rapidly-varying electric dipole moment \cite{Schrodinger1930}. Also, the electron instantaneous velocity in the Dirac theory is the velocity of light, and oscillates rapidly around the mean motion. Schr\"odinger called the electron oscillatory motion of the Dirac theory ``zitterbewegung.''  The zitterbewegung may be regarded as a circular motion at the speed of light around the direction of the electron spin, giving rise to the electron intrinsic magnetic moment \cite{Huang1952}. The phase of the zitterbewegung motion has been proposed \cite{Hestenes1990oc1} to provide a physical interpretation for the complex phase factor of the Dirac wave function. As will be shown, representation of charged particles as circulating point charges leads to a classical expectation for the magnetic spin-spin interaction between similar (i.e., equal mass) particles that under certain circumstances is remarkably similar to the ground state quantum force in hydrogenic atoms.

The thesis of this paper, that the quantum force of Bohm's theory may plausibly have an origin in classical electrodynamics, relies on the determination by Rivas \cite{Rivas} of the electromagnetic structure of an electron consisting of a luminally circulating classical point charge. Rivas has shown that if the velocity of the circulatory motion is the speed of light, then the average electric field is purely radial, isotropic (except in the plane of the motion, where it is undefined) and, sufficently far from the center of the motion, is consistent in magnitude with the field of a stationary charge of equal magnitude, falling off inverse squarely with distance. Rivas also shows, for radius of motion of half the reduced Compton wavelength and where it is defined, at sufficient distance the time averaged magnetic acceleration field of the circulating charge is equal to the magnetic field of an ideal dipole with moment of half the Bohr magneton. The interaction Lagrangian for two spin 1/2 Dirac particles \cite{RivasY7} has also been obtained based on this approach. 

In this paper the analysis of the electromagnetic field structure of the luminally circulating charge by Rivas is extended only slightly to consider the magnetic interaction between two equal mass particles each consisting of a luminally circulating charge.  However, a simpler approach than Lagrangian mechanics will be used to characterize the interaction.  Rather than attempting to apply the interaction Lagrangian, it will simply be assessed what is the magnetic force which one luminally circulating charge exerts upon another charge undergoing similar motion, while neglecting the influence of the second circulating charge upon the first. One of the particles is thus taken as the source of an electromagnetic field acting on a second, test, particle.  As it has already been shown by Rivas that the average of the electric acceleration field is Coulomb-like in the far field, except in the plane of the motion where it is undefined, it is a small step to see that the magnetic force on a second luminally circulating charge will also follow an inverse-square depedence on interparticle separation in the far field.  The magnetic field strength is equal (in Gaussian units) to the electric field strength, while the magnetic force magnitude is bounded by the magnetic field strength times the test particle velocity ratio to the speed of light.  For luminally circulating charges, the magnetic force in the far field is thus similar in distance dependence to the Coulomb force between static charges. 

For further simplification it will also be assumed that the radii of the circular motions of the two particles are equal, and that the particles are widely separated compared to the radius of the motions. Also, the angular momenta of the circulatory motions of the two particles, representing the spin vectors, will be assumed aligned. Because the test particle is moving at the speed of light in a field that is of the same mean magnitude as a Coulomb field of a charge of equal value to the field source charge, the magnetic interaction has nominally the same strength as the Coulomb interaction.  However, it is modulated by both the relative orientation and the relative phase of the circular motions. Furthermore, the relevant phase difference varies with interparticle separation, because it is between the test particle present-time phase and the field source particle retarded-time phase.

\section{Retarded Electromagnetic Field of a Relativistically Circulating Charge}

In this section the electromagnetic field of a luminally-circulating point charge is evaluated explicitly, so that in the next section the magnetic force on the luminally-circulating test charge can be evaluated. Sections \ref{ssectzbwEfield} and \ref{ssectzbwBfield} here reproduce Rivas's work in \cite{Rivas} with notation tailored to support the derivation of the magnetic force of one luminally-circulating charge upon another of Section \ref{sectzbwmagradforce}. 

The electromagnetic field due to a point charge in arbitrary motion can be determined from the Li\'enard-Wiechert potentials. The electric and magnetic fields, \(\mbox{\boldmath$E$}\) and \(\mbox{\boldmath$B$}\), at a field point  \(\mbox{\boldmath$r$}\) and time \(t\) for an arbitrarily moving field-source charge \(q_s\) obtained from the retarded Li\'enard-Wiechert potentials may be expressed (in Gaussian units) as  \cite{jcksn:classelec}

\begin{widetext}

\begin{equation}
\mbox{\boldmath$E$}(\mbox{\boldmath$r$},t) = q_s \left[ \frac{\mbox{\boldmath$n$}-\mbox{\boldmath$\beta$}}{\gamma^2 \left( 1 - \mbox{\boldmath$\beta$} \cdot \mbox{\boldmath$n$}\right)^3 R^2 }\right]_{\textnormal{{\small ret}}} + \frac{q_s}{c} \left[ \frac{\mbox{\boldmath$n$} \times \left((\mbox{\boldmath$n$}-\mbox{\boldmath$\beta$})\times \dot{\mbox{\boldmath$\beta$}}\right)}{\left(1 - \mbox{\boldmath$\beta$} \cdot \mbox{\boldmath$n$}\right)^3 R }\right]_{\textnormal{{\small ret}}},
\label{LW_E_field}
\end{equation}

\end{widetext}

\begin{equation}
\mbox{\boldmath$B$}(\mbox{\boldmath$r$},t) = \left[ \mbox{\boldmath$n$} \times \mbox{\boldmath$E$}\right]_{\textnormal{{\small ret}}}, 
\label{LW_B_field}
\end{equation}

where \(\mbox{\boldmath$n$}\) is a unit vector in the direction of the field point from the position of the field-source charge at the retarded time \(t_r = t - R/c\), with \(R\) the magnitude of the displacement from the charge position at the retarded time to the field point \(\mbox{\boldmath$r$}\) at time \(t\).   With the field-source charge velocity \(\mbox{\boldmath$v$}\), \(\mbox{\boldmath$\beta$} = \mbox{\boldmath$v$}/c\), and \( \gamma \equiv (1 - \beta^2)^{-1/2}\).  The subscript ``ret'' indicates that quantities in the brackets are evaluated at the retarded time.  The overdot represents differentiation with respect to \(t\), so \(\dot{\mbox{\boldmath$\beta$}} = \mbox{\boldmath$a$}/c \) where \(\mbox{\boldmath$a$} = a \hat{\mbox{\boldmath$a$}}\) is the acceleration.

The first term on the right hand side of Eq. (\ref{LW_E_field}), often referred to as the electric velocity field, vanishes in the limit of \(\beta\) approaching unity, except if \(\mbox{\boldmath$\beta$} = \mbox{\boldmath$n$}\), when it is undefined due to vanishing of the inverse factor of \((1 - \mbox{\boldmath$\beta$} \cdot \mbox{\boldmath$n$})^3  \). The second term on the right hand side of Eq. (\ref{LW_E_field}), also undefined when \(\mbox{\boldmath$\beta$} = \mbox{\boldmath$n$}\), can be referred to as the electric acceleration field, or as the radiative field owing to its inverse \(R\) rather than inverse \(R^2\) dependence.  However, the acceleration field becomes a solution to the electromagnetic wave equation only in the far field. It can also give rise to more rapidly decaying field terms that are dynamically relevant in the near field.

\subsection{Electric Field of a Classical Zitterbewegung Particle}
\label{ssectzbwEfield}

To determine the magnetic force between two relativistically circulating charges, the Li\'enard-Wiechert retarded electric acceleration field is first evaluated for the source charge undergoing circular motion. (To avoid the singular condition where \(\mbox{\boldmath$\beta$} = \mbox{\boldmath$n$}\), it is sufficient assume that the test charge center of circular motion is away from the plane of circular motion of the source charge.  This also ensures that the electric velocity field vanishes, for the luminal field-source charge.) Next, the magnetic field can be evaluated using Eq. (\ref{LW_B_field}), and the magnetic force on the luminally-circulating test charge follows straightforwardly.  

For brevity, and after \cite{HestenesY10}, it will be convenient to refer to the luminal circular motion of the point charge as the ``zitter'' motion, and to a charge undergoing such motion as a zitter particle.  In spite of the luminal circular motion of its charge, the zitter particle is considered stationary when its center of motion is fixed.

For the stationary zitter particle, it is useful to suppose that the center of the zitter motion is the origin of a Cartesian coordinate system, so that \(\mbox{\boldmath$r$}\) of Eqs. (\ref{LW_E_field}) and (\ref{LW_B_field}) can represent the vector displacement from the center of the zitter motion of the source charge to the field point. Then if \(\mbox{\boldmath$k$}\) is the vector displacement from the center to the charge instantaneous position,  \(\mbox{\boldmath$k$} = \lambdabar_0 \hat{\mbox{\boldmath$k$}}\) where \(\hat{\mbox{\boldmath$k$}}\) is a unit vector and \(\lambdabar_0\) is the radius of the circular motion.  With angular momentum due to the zitter motion given as \(L = \lambdabar_0 \gamma m v \) with \(v = c\), taking \(\gamma m = m_e\) where \(m_e\) is the observed electron mass, and \(\lambdabar_0  = \hbar / 2 m_e c\), obtains that the electron intrinsic angular momentum is \(\hbar/2\). The diameter of the circulatory motion that has angular momentum \(\hbar/2\) is thus one reduced Compton wavelength.  Also, for the stationary zitter particle, the acceleration magnitude \(a = c|\dot{\mbox{\boldmath$\beta$}}| = c^2/\lambdabar_0\). Defining the plane of the motion as the \(x y\) or \(z=0\) plane of a right-handed Cartesian coordinate system \((x,y,z)\), then for the case of counterclockwise zitter motion,  \(\hat{\mbox{\boldmath$k$}} = [\cos \omega t, \sin \omega t, 0]\) with \(\omega = c/\lambdabar_0\).

If  \(\mbox{\boldmath$R$} = \mbox{\boldmath$r$} - \mbox{\boldmath$k$} = r\hat{\mbox{\boldmath$r$}} - \lambdabar_0\hat{\mbox{\boldmath$k$}} \) is the displacement from the instantaneous charge position to the field point, then \(R \equiv \left|\mbox{\boldmath$R$}\right| = (r^2 - 2\lambdabar_0\mbox{\boldmath$r$} \cdot \hat{\mbox{\boldmath$k$}}  + {\lambdabar_0}^2 )^{1/2} = r(1 - 2\epsilon\hat{\mbox{\boldmath$r$}} \cdot \hat{\mbox{\boldmath$k$}} + {\epsilon}^2 )^{1/2}\) with \(\epsilon = \lambdabar_0/r\), and \(\mbox{\boldmath$n$} \equiv \mbox{\boldmath$R$}/R\) becomes

\begin{equation}
\mbox{\boldmath$n$} = \frac{\hat{\mbox{\boldmath$r$}} - \epsilon \hat{\mbox{\boldmath$k$}}}{ (1 - 2\epsilon\hat{\mbox{\boldmath$r$}} \cdot \hat{\mbox{\boldmath$k$}} + {\epsilon}^2 )^{1/2}}.
\label{nvec}
\end{equation}

The field-source charge acceleration is toward the center of the circular motion, so \(\hat{\mbox{\boldmath$a$}} = -\hat{\mbox{\boldmath$k$}}\), and for  counter-clockwise circular motion \( \mbox{\boldmath$\beta$} \times \hat{\mbox{\boldmath$a$}} = \hat{\mbox{\boldmath$z$}} \), and so

\begin{equation}
(\mbox{\boldmath$n$}-\mbox{\boldmath$\beta$}) \times \hat{\mbox{\boldmath$a$}} = -\frac{\hat{\mbox{\boldmath$r$}} \times \hat{\mbox{\boldmath$k$}}}{ (1 - 2\epsilon\hat{\mbox{\boldmath$r$}} \cdot \hat{\mbox{\boldmath$k$}} + {\epsilon}^2 )^{1/2}  } - \hat{\mbox{\boldmath$z$}}.
\label{nmbxahat}
\end{equation}

Also for  counter-clockwise circular motion,  \(\hat{\mbox{\boldmath$k$}} \times  \hat{\mbox{\boldmath$z$}} = -\mbox{\boldmath$\beta$}\), and cross-multiplying Eq. (\ref{nmbxahat}) by Eq. (\ref{nvec}) on the left yields

\begin{widetext}

\begin{equation}
\mbox{\boldmath$n$} \times \left(\left(\mbox{\boldmath$n$}-\mbox{\boldmath$\beta$}\right) \times \hat{\mbox{\boldmath$a$}}\right) = -\frac{\left(\hat{\mbox{\boldmath$r$}} \cdot \hat{\mbox{\boldmath$k$}}\right)\left(\hat{\mbox{\boldmath$r$}} - \epsilon  \hat{\mbox{\boldmath$k$}}\right)  - \hat{\mbox{\boldmath$k$}} - \epsilon\hat{\mbox{\boldmath$r$}}}{1 - 2\epsilon\hat{\mbox{\boldmath$r$}} \cdot \hat{\mbox{\boldmath$k$}} + {\epsilon}^2  } - \frac{\hat{\mbox{\boldmath$r$}} \times\hat{\mbox{\boldmath$z$}} +  \epsilon \mbox{\boldmath$\beta$}}{ (1 - 2\epsilon\hat{\mbox{\boldmath$r$}} \cdot \hat{\mbox{\boldmath$k$}} + {\epsilon}^2 )^{1/2}},
\label{nxnmbxahat}
\end{equation}

\end{widetext}

where the vector identity \(\mbox{\boldmath$a$} \times \left(\mbox{\boldmath$b$} \times \mbox{\boldmath$c$} \right) =  \left(\mbox{\boldmath$a$} \cdot \mbox{\boldmath$c$} \right)\mbox{\boldmath$b$} - \left(\mbox{\boldmath$a$} \cdot \mbox{\boldmath$b$} \right)\mbox{\boldmath$c$}\) has been used.

Recalling that \(\dot{\mbox{\boldmath$\beta$}} = a\hat{\mbox{\boldmath$a$}}/c\), with \(a = c^2 / \lambdabar_0\) for the stationary zitter particle and that \(R = r(1 - 2\epsilon\hat{\mbox{\boldmath$r$}} \cdot \hat{\mbox{\boldmath$k$}} + {\epsilon}^2 )^{1/2}\), the electric acceleration field term from Eq. (\ref{LW_E_field}) can now be written in terms of the distance \(r\) from the zitter particle center of motion (rather than in terms of the distance \(R\) from the luminally-circulating charge position) as

\begin{widetext}

\begin{equation}
\mbox{\boldmath$E$}_{\textnormal{{\scriptsize acc}}}(\mbox{\boldmath$r$},t) = \frac{-q_s}{\lambdabar_0\left(1 - \mbox{\boldmath$\beta$} \cdot \mbox{\boldmath$n$}\right)^3 r} \left[\frac{\left(\hat{\mbox{\boldmath$r$}} \cdot \hat{\mbox{\boldmath$k$}}\right)\left(\hat{\mbox{\boldmath$r$}} - \epsilon  \hat{\mbox{\boldmath$k$}}\right)  - \hat{\mbox{\boldmath$k$}} - \epsilon\hat{\mbox{\boldmath$r$}}}{(1 - 2\epsilon\hat{\mbox{\boldmath$r$}} \cdot \hat{\mbox{\boldmath$k$}} + {\epsilon}^2 )^{3/2} } + \frac{\hat{\mbox{\boldmath$r$}} \times\hat{\mbox{\boldmath$z$}} +  \epsilon \mbox{\boldmath$\beta$}}{1 - 2\epsilon\hat{\mbox{\boldmath$r$}} \cdot \hat{\mbox{\boldmath$k$}} + {\epsilon}^2 }\right],
\label{E_acc}
\end{equation}

\end{widetext}

where all quantities on the right hand side are evaluated at the retarded time.  

Although the electric acceleration field due to the zitter particle has an explicit inverse dependence on the distance \(r\), the factors of \(\epsilon = \lambdabar_0/r\) give dependencies of inverse higher powers of \(r\), including, but not limited to, inverse \(r^2\).     

To see clearly the inverse square of distance character of the time-averaged electric acceleration field intensity, it is useful to consider the behavior of the field on the z-axis, i.e., perpendicular to the plane of the zitter motion of the source charge.  When \( \hat{\mbox{\boldmath$r$}} = \hat{\mbox{\boldmath$z$}}\), then \( \hat{\mbox{\boldmath$r$}} \cdot \hat{\mbox{\boldmath$k$}} = \hat{\mbox{\boldmath$r$}} \times\hat{\mbox{\boldmath$z$}} = 0\),  \(\mbox{\boldmath$\beta$} \cdot \mbox{\boldmath$n$} = 0\), and so

\begin{equation}
\mbox{\boldmath$E$}_{\textnormal{{\scriptsize acc}}}(\mbox{\boldmath$r$}=z\hat{\mbox{\boldmath$z$}},t) = \frac{q_s}{\lambdabar_0 z} \left[\frac{ \hat{\mbox{\boldmath$k$}} + \epsilon\hat{\mbox{\boldmath$z$}}}{(1  + {\epsilon}^2 )^{3/2} } - \frac{\epsilon \mbox{\boldmath$\beta$}}{1  + {\epsilon}^2}\right].
\end{equation}

The unit vectors \( \hat{\mbox{\boldmath$k$}}\) and \(\mbox{\boldmath$\beta$}\) average to zero over a cycle of the zitter motion, and on the z-axis \(\epsilon = \lambdabar_0/z\).  The average electric acceleration field on the z-axis is thus

\begin{equation}
<\mbox{\boldmath$E$}_{\textnormal{{\scriptsize acc}}}(\mbox{\boldmath$r$}=z\hat{\mbox{\boldmath$z$}})> = \frac{q_s}{z^2} \left[\frac{ \hat{\mbox{\boldmath$z$}}}{(1  + {\epsilon}^2 )^{3/2} } \right].
\label{Rivas_KTSP_6.3}
\end{equation}

As \(z\) becomes large, \(1+\epsilon^2\) approaches unity, and so the long distance behavior of the time-averaged electric acceleration field of the fixed zitter particle, perpendicular to the plane of the zitter motion, is indistinguishable from the electric field of a static charge of the same magnitude.  (Eq. (\ref{Rivas_KTSP_6.3}) here is equivalent to Eq. (6.3) of \cite{Rivas_oc}.)

Rivas \cite{Rivas_oc} has numerically integrated the electric acceleration field of the zitter particle to find the average at points off the z-axis, and at near as well as distant separation from center of the zitter motion.  The average field is purely radial and isotropic at distances large compared to the zitter radius at all points off the \(z=0\) plane. On the \(z=0\) plane, \(\left(1 - \mbox{\boldmath$\beta$} \cdot \mbox{\boldmath$n$}\right)\) can vanish and the electric field average cannot be calculated.

\subsection{Magnetic Field of a Classical Zitterbewegung Particle}
\label{ssectzbwBfield}

The magnetic field due to the zitter particle charge acceleration is evaluated using Eqs. (\ref{LW_B_field}) and (\ref{E_acc}), but it is convenient to use the intermediate result of Eq. (\ref{nxnmbxahat}).  Taking the cross product of Eq. (\ref{nxnmbxahat}) with \(\mbox{\boldmath$n$}\) (as given by Eq. (\ref{nvec})) on the left obtains that

\begin{eqnarray}
\mbox{\boldmath$n$} \times \left[\mbox{\boldmath$n$} \times \left(\left(\mbox{\boldmath$n$}-\mbox{\boldmath$\beta$}\right) \times \hat{\mbox{\boldmath$a$}}\right)\right] = \frac{ (1 - \epsilon^2)\hat{\mbox{\boldmath$r$}} \times \hat{\mbox{\boldmath$k$}}}{(1 - 2\epsilon\hat{\mbox{\boldmath$r$}} \cdot \hat{\mbox{\boldmath$k$}} + {\epsilon}^2 )^{3/2} } \nonumber \\  - \frac{\left(\hat{\mbox{\boldmath$r$}} - \epsilon\hat{\mbox{\boldmath$k$}}\right) \times\left(\hat{\mbox{\boldmath$r$}} \times\hat{\mbox{\boldmath$z$}}\right) +  \epsilon \left(\hat{\mbox{\boldmath$r$}} - \epsilon\hat{\mbox{\boldmath$k$}}\right) \times \mbox{\boldmath$\beta$}}{(1 - 2\epsilon\hat{\mbox{\boldmath$r$}} \cdot \hat{\mbox{\boldmath$k$}} + {\epsilon}^2 )},
\end{eqnarray}

and with the vector identity \(\mbox{\boldmath$a$} \times \left(\mbox{\boldmath$b$} \times \mbox{\boldmath$c$} \right) =  \left(\mbox{\boldmath$a$} \cdot \mbox{\boldmath$c$} \right)\mbox{\boldmath$b$} - \left(\mbox{\boldmath$a$} \cdot \mbox{\boldmath$b$} \right)\mbox{\boldmath$c$}\), 

\begin{widetext}

\begin{eqnarray}
\mbox{\boldmath$n$} \times \left[\mbox{\boldmath$n$} \times \left(\left(\mbox{\boldmath$n$}-\mbox{\boldmath$\beta$}\right) \times \hat{\mbox{\boldmath$a$}}\right)\right] = \frac{(1 - \epsilon^2)\hat{\mbox{\boldmath$r$}} \times \hat{\mbox{\boldmath$k$}}}{(1 - 2\epsilon\hat{\mbox{\boldmath$r$}} \cdot \hat{\mbox{\boldmath$k$}} + {\epsilon}^2 )^{3/2}}  - \frac{\left(\hat{\mbox{\boldmath$r$}} \cdot \hat{\mbox{\boldmath$z$}} \right)\hat{\mbox{\boldmath$r$}} - \hat{\mbox{\boldmath$z$}}  - \epsilon\left( \left(\hat{\mbox{\boldmath$k$}} \cdot \hat{\mbox{\boldmath$z$}} \right)\hat{\mbox{\boldmath$r$}} - \left(\hat{\mbox{\boldmath$k$}} \cdot \hat{\mbox{\boldmath$r$}} \right)\hat{\mbox{\boldmath$z$}}\right) +  \epsilon \left(\hat{\mbox{\boldmath$r$}} - \epsilon\hat{\mbox{\boldmath$k$}}\right) \times \mbox{\boldmath$\beta$}}{(1 - 2\epsilon\hat{\mbox{\boldmath$r$}} \cdot \hat{\mbox{\boldmath$k$}} + {\epsilon}^2 )}.
\label{btxnxnmbxahat}
\end{eqnarray}

With  \(a = c^2/\lambdabar_0\), \(\dot{\mbox{\boldmath$\beta$}} = a\hat{\mbox{\boldmath$a$}}/c\), and \(R = r(1 - 2\epsilon\hat{\mbox{\boldmath$r$}} \cdot \hat{\mbox{\boldmath$k$}} + {\epsilon}^2 )^{1/2}\), the magnetic acceleration field of the zitterbewegung particle can now be found from Eqs. (\ref{LW_B_field}) and (\ref{E_acc}) to be

\begin{eqnarray}
\mbox{\boldmath$B$}_{\textnormal{{\scriptsize acc}}} =  \frac{q_s}{\lambdabar_0\left(1 - \mbox{\boldmath$\beta$} \cdot \mbox{\boldmath$n$}\right)^3 r}\left[\frac{(1 - \epsilon^2)\hat{\mbox{\boldmath$r$}} \times \hat{\mbox{\boldmath$k$}}}{(1 - 2\epsilon\hat{\mbox{\boldmath$r$}} \cdot \hat{\mbox{\boldmath$k$}} + {\epsilon}^2 )^{2}} \right. -  \left. \frac{\left(\hat{\mbox{\boldmath$r$}} \cdot \hat{\mbox{\boldmath$z$}} \right)\hat{\mbox{\boldmath$r$}} - \hat{\mbox{\boldmath$z$}}  - \epsilon\left( \left(\hat{\mbox{\boldmath$k$}} \cdot \hat{\mbox{\boldmath$z$}} \right)\hat{\mbox{\boldmath$r$}} - \left(\hat{\mbox{\boldmath$k$}} \cdot \hat{\mbox{\boldmath$r$}} \right)\hat{\mbox{\boldmath$z$}}\right) +  \epsilon \left(\hat{\mbox{\boldmath$r$}} - \epsilon\hat{\mbox{\boldmath$k$}}\right) \times \mbox{\boldmath$\beta$}}{(1 - 2\epsilon\hat{\mbox{\boldmath$r$}} \cdot \hat{\mbox{\boldmath$k$}} + {\epsilon}^2 )^{3/2}}\right],
\label{zbw_B_accel}
\end{eqnarray}

\end{widetext}

where the right-hand side is evaluated at the retarded time.

Since for luminal charge velocity the electric velocity field vanishes everywhere that it is defined, and the acceleration field is undefined under the same conditions as the velocity field, Eq. (\ref{zbw_B_accel}), where it is well defined, describes the complete magnetic field of the zitter particle. 

\section{Radially-Directed Inverse-Square Law Magnetic Force Between Dirac Particles}
\label{sectzbwmagradforce}

The force on a test charge \(q_t\) moving with velocity \(\mbox{\boldmath$\beta$}_t\)  in a magnetic field \(\mbox{\boldmath$B$}\)  is  \(\mbox{\boldmath$F$} =  q_t \mbox{\boldmath$\beta$}_t \times \mbox{\boldmath$B$} \). The magnetic force on a test particle in arbitrary motion, due to the acceleration field of a source particle undergoing the circular zitter motion of radius \(\lambdabar_0\) can be evaluated by taking a cross product of Eq. (\ref{zbw_B_accel}) with \(\mbox{\boldmath$\beta$}_t\), and using again the vector identity \(\mbox{\boldmath$a$} \times \left(\mbox{\boldmath$b$} \times \mbox{\boldmath$c$} \right) =  \left(\mbox{\boldmath$a$} \cdot \mbox{\boldmath$c$} \right)\mbox{\boldmath$b$} - \left(\mbox{\boldmath$a$} \cdot \mbox{\boldmath$b$} \right)\mbox{\boldmath$c$}\), to obtain that 

\begin{widetext}

\begin{eqnarray}
\mbox{\boldmath$F$} = \frac{q_t q_s}{\lambdabar_0\left(1 - \mbox{\boldmath$\beta$} \cdot \mbox{\boldmath$n$}\right)^3 r} \left[ \frac{(1 - \epsilon^2)\left[\left(\mbox{\boldmath$\beta$}_t \cdot \hat{\mbox{\boldmath$k$}} \right)\hat{\mbox{\boldmath$r$}} - \left(\mbox{\boldmath$\beta$}_t \cdot \hat{\mbox{\boldmath$r$}} \right)\hat{\mbox{\boldmath$k$}}\right]}{(1 - 2\epsilon\hat{\mbox{\boldmath$r$}} \cdot \hat{\mbox{\boldmath$k$}} + {\epsilon}^2 )^{2}}  \right. \nonumber \\  - \frac{\left(\hat{\mbox{\boldmath$r$}} \cdot \hat{\mbox{\boldmath$z$}} \right)\mbox{\boldmath$\beta$}_t \times\hat{\mbox{\boldmath$r$}} - \mbox{\boldmath$\beta$}_t \times\hat{\mbox{\boldmath$z$}}  + \epsilon \left(\hat{\mbox{\boldmath$k$}} \cdot \hat{\mbox{\boldmath$r$}} \right)\mbox{\boldmath$\beta$}_t \times\hat{\mbox{\boldmath$z$}} +  \epsilon\left[\left(\mbox{\boldmath$\beta$}_t \cdot \mbox{\boldmath$\beta$} \right)\hat{\mbox{\boldmath$r$}} - \left(\mbox{\boldmath$\beta$}_t \cdot \hat{\mbox{\boldmath$r$}} \right)\mbox{\boldmath$\beta$}\right]}{(1 - 2\epsilon\hat{\mbox{\boldmath$r$}} \cdot \hat{\mbox{\boldmath$k$}} + {\epsilon}^2 )^{3/2}}  + \left. \frac{\epsilon^2\left[ \left(\mbox{\boldmath$\beta$}_t \cdot \mbox{\boldmath$\beta$} \right)\hat{\mbox{\boldmath$k$}} - \left(\mbox{\boldmath$\beta$}_t \cdot \hat{\mbox{\boldmath$k$}} \right)\mbox{\boldmath$\beta$}\right]}{(1 - 2\epsilon\hat{\mbox{\boldmath$r$}} \cdot \hat{\mbox{\boldmath$k$}} + {\epsilon}^2 )^{3/2}}   \right],
\label{gen_zbw_mag_force}
\end{eqnarray}

\end{widetext}

where the right-hand side is evaluated at the retarded time, except  \(\mbox{\boldmath$\beta$}_t\). It is apparent that the magnetic force will have a radially-directed term ({ \em i.e.}, the term \(\epsilon\left(\mbox{\boldmath$\beta$}_t \cdot \mbox{\boldmath$\beta$}\right)\hat{\mbox{\boldmath$r$}}\) in the numerator of the second fraction in the large square brackets of Eq. (\ref{gen_zbw_mag_force})), 

\begin{widetext}

\begin{eqnarray}
\mbox{\boldmath$F$}_r =  -\frac{q_t q_s}{\lambdabar_0} \mbox{\boldmath$\beta$}_t \cdot\left[\frac{  \epsilon \mbox{\boldmath$\beta$}\hat{\mbox{\boldmath$r$}}}{\left(1 - \mbox{\boldmath$\beta$} \cdot \mbox{\boldmath$n$}\right)^3 r (1 - 2\epsilon\hat{\mbox{\boldmath$r$}} \cdot \hat{\mbox{\boldmath$k$}} + \epsilon^2)^{3/2}}\right]_{\textnormal{{\scriptsize ret}}} =  -\frac{q_t q_s \hat{\mbox{\boldmath$r$}}}{r^2} \mbox{\boldmath$\beta$}_t \cdot\left[\frac{\mbox{\boldmath$\beta$}}{\left(1 - \mbox{\boldmath$\beta$} \cdot \mbox{\boldmath$n$}\right)^3 (1 - 2\epsilon\hat{\mbox{\boldmath$r$}} \cdot \hat{\mbox{\boldmath$k$}} + {\epsilon}^2 )^{3/2}}\right]_{\textnormal{{\scriptsize ret}}}, 
\label{MagForceRadialTerm}
\end{eqnarray}

\end{widetext}

where in the second equality \(\hat{\mbox{\boldmath$r$}}\) and \(r\) have been taken outside the brackets indicating retardation, because the center of motion of the zitterbewegung field-source particle is assumed to be stationary, and it has been used again that \(\epsilon = \lambdabar_0/r\). For large enough interparticle separation, \(\epsilon << 1\) and so

\begin{equation}
\mbox{\boldmath$F$}_r \approx  \frac{-q_t q_s \mbox{\boldmath$\beta$}_t \cdot   \mbox{\boldmath$\beta$}_{s,{\textnormal{{\scriptsize ret}}}} \hat{\mbox{\boldmath$r$}}}{\left(1 - \mbox{\boldmath$\beta$} \cdot \mbox{\boldmath$n$}\right)^3 r^2},
\label{radial_force_term}
\end{equation}

where in the numerator \(\mbox{\boldmath$\beta$}\) is rewritten as \(\mbox{\boldmath$\beta$}_{s,{\textnormal{{\scriptsize ret}}}}\) to emphasize its association with the source particle and at the retarded time.  The term given by Eq. (\ref{radial_force_term}), that contributes to the magnetic force caused by a stationary zitter particle on another zitter particle, differs from the Coulomb force caused by a stationary charge on another classical point charge by the factor \(-\mbox{\boldmath$\beta$}_t \cdot \mbox{\boldmath$\beta$}_{s,{\textnormal{{\scriptsize ret}}}}\)  and the inverse factor of \((1 - \mbox{\boldmath$\beta$} \cdot \mbox{\boldmath$n$})^3\).  The inverse factor of \((1 - \mbox{\boldmath$\beta$} \cdot \mbox{\boldmath$n$})^3\)  modulates the magnitude of the radial magnetic force at the frequency of the zitter motion, but averages to unity over a zitter cycle.

To make a preliminary assessment of how the magnetic force radial term given by Eq. (\ref{radial_force_term}) will affect the motion of a zitterbewegung test particle, the test particle may be supposed to be executing a circular motion with the same radius \(\lambdabar_0\) as that of the field source particle, and also at the speed of light so that \(\beta_t = 1\). (The inverse \((1 - \mbox{\boldmath$\beta$} \cdot \mbox{\boldmath$n$})^3\) factor will be disregarded here, supposing that its physical effects average out or that the test particle is located close enough to the \(z\)-axis so that \(\mbox{\boldmath$\beta$} \cdot \mbox{\boldmath$n$}\) is negligibly small.) For simplicity, suppose both particles have their zitter motion in the x-y plane and in the same direction. Also, while strictly \(\hat{\mbox{\boldmath$r$}}\) is directed from the source particle center of motion toward a field point moving with the test particle, for interparticle separation large compared to \(\lambdabar_0\), {\em i.e.,} large compared to half a reduced Compton wavelength, it is reasonable to disregard the change in position of the test particle as it moves in its zitter orbit, and consider only its velocity direction time dependence. Then,

\begin{equation}
 \mbox{\boldmath$\beta$}_t \cdot   \mbox{\boldmath$\beta$}_{s,{\textnormal{{\scriptsize ret}}}} = [-\sin\omega t,\cos\omega t,0] \cdot [-\sin(\omega t + \delta\phi),\cos(\omega t + \delta\phi),0]  
\end{equation}

where \(\delta \phi\) is the phase difference of the test particle present-time zitter motion to the retarded-time source particle zitter motion. Carrying out the multiplication obtains that

\begin{equation}
 \mbox{\boldmath$\beta$}_t \cdot   \mbox{\boldmath$\beta$}_{s,{\textnormal{{\scriptsize ret}}}} = \cos\delta\phi.
\end{equation}

Thus, depending on the zitter phase difference between the particles, for aligned spins, the magnetic force may effectively either cancel or double the Coulomb force between the particles. 

In the case of the test particle circulatory motion opposite to that of the field source particle,

\begin{equation}
 \mbox{\boldmath$\beta$}_t \cdot   \mbox{\boldmath$\beta$}_{s,{\textnormal{{\scriptsize ret}}}} = -(\sin^2\omega t - \cos^2\omega t)\cos\delta\phi-\sin2\omega t\sin\delta\phi,
\end{equation}

which averages to zero over a zitter cycle.  However, the magnetic force inverse square law radial component may be found to nonetheless have an influence on the motion, when the average is more properly carried out including the change in position of the test charge over its zitter orbit.  Also, there are other inverse-square law magnetic force terms in Eq. (\ref{gen_zbw_mag_force}) which have yet to be considered.

For aligned zitter motions, it is straightforward to assess how the change in the source particle zitter phase induced by retardation might influence the test particle motion under the combination of electric and magnetic forces.  The phase of the test particle zitter motion can be expressed as \(\phi_t = \phi_{t0} + \omega t\), where \( \phi_{t0} \equiv \phi_t(t=0) \) is the test particle zitter phase at \(t=0\).   Similarly, for aligned spins, the phase of the source particle zitter motion can be expressed as \(\phi_s = \phi_{s0} + \omega t\), where \( \phi_{s0} \) is the source particle zitter phase at \(t=0\). Then the source particle phase at the retarded time is

\begin{equation}
\phi_s(t_r) \equiv \phi_{s,{\textnormal{{\scriptsize ret}}}} = \phi_{s0} + \omega\left(t -  r_{{\textnormal{{\scriptsize ret}}}}/c\right) = \phi_s(t) -  r_{{\textnormal{{\scriptsize ret}}}}/\lambdabar_0,
\end{equation}

where \( r_{{\textnormal{{\scriptsize ret}}}} \equiv r(t_r)\). Thus, if \(\delta\phi_0 \equiv \phi_{t0} - \phi_{s0}\) is the phase difference between the zitter motions neglecting retardation, then the phase difference \(\delta\phi_{{\textnormal{{\scriptsize ret}}}} \equiv \phi_t - \phi_{s,{\textnormal{{\scriptsize ret}}}}\) accounting for retardation is

\begin{equation}
\delta\phi_{{\textnormal{{\scriptsize ret}}}}(t) = \delta\phi_0 + \frac{r_{{\textnormal{{\scriptsize ret}}}}(t)}{\lambdabar_0}.
\end{equation}

Therefore, over the course of an interparticle distance change by an amount equal to \(\pi\) times the reduced Compton wavelength \(2 \lambdabar_0\) of the particles, the magnetic force radial term of Eq. (\ref{MagForceRadialTerm}) will undergo a full cycle of sinusoidal modulation. Also, because the radial magnetic force can have a magnitude up to as large as the Coulomb force between two static charges of the same magnitude, it can be expected to significantly influence the motion. However, it's difficult to envision how the magnetic force derived here can be directly linked to the Bohmian quantum force, since a change in separation of only one-half the Compton wavelength will invert the sign of the force.  Since the Bohr radius can be expressed as the Compton wavelength divided by the fine-structure constant, this corresponds to a change in separation of only about \(\pi/(2(137))\) times the Bohr radius.  The Bohmian quantum force, as shown in the appendix, does not undergo such sign reversals in the ground state.

\section{Time Dilation and Relationship to the De Broglie Wavelength}
\label{secttimedilation}

Up to this point the inverse-square-law magnetic force between zitter particles has been calculated assuming that the centers of zitter motion of the source and test particles are relatively stationary. Since it was noted that the magnetic force between stationary zitter particles varies sinusoidally with the separation between the particles, it is naturally of interest to consider what is the effect of relative motion between the zitter particles. The effect on the relative zitter phase due to relative motion can be straightforwardly taken into account. The effect of relativistic time dilation on the zitter frequency of a moving zitter particle must also be taken into account.  For the present any physical consequences are yet to be worked out, but it seems worth noting that the wavelength associated with the beat frequency between the rest frame and time dilated zitter frequencies can be equated with the de Broglie wavelength in the limit of small relative velocity between the zitter particles.

Suppose that the test zitter particle is uniformly translating with speed \(v = c |\bar{\mbox{\boldmath$\beta$}_t}|\), where the overbar indicates an average over a test particle zitter period, in the inertial reference frame where the  field-source zitter particle is stationary. Then, by the Lorentz transformation, the phase of the test particle zitter motion as observed in the source zitter particle rest frame can be written as \cite{BaylisY7} \(\omega \tau = \gamma \omega( t - \mbox{\boldmath$\upsilon$} \cdot \mbox{\boldmath$r$}/c^2)\), where \(\tau\) is the time coordinate in the test zitter particle rest frame, \(\gamma = (1 - (v/c)^2)^{-1/2}\), and \(\mbox{\boldmath$r$}\) here is the displacement from the source zitter particle to the test zitter particle. The modulating factor on the inverse-square radial magnetic force in the numerator of the right-hand side of Eq. (\ref{radial_force_term}) becomes

\begin{widetext}

\begin{equation}
 \mbox{\boldmath$\beta$}_t \cdot   \mbox{\boldmath$\beta$}_{s,{\textnormal{{\scriptsize ret}}}} = [-\sin\gamma \omega( t - \mbox{\boldmath$\upsilon$} \cdot \mbox{\boldmath$r$}/c^2) ,\cos \gamma \omega( t - \mbox{\boldmath$\upsilon$} \cdot \mbox{\boldmath$r$}/c^2) ,0] \cdot [-\sin(\omega t_r + \delta\phi),\cos(\omega t_r + \delta\phi_0),0],  
\label{first_correction} 
\end{equation}

\end{widetext}

where \(\delta\phi_0\) here is a constant to account for the phase difference between the source and test particle zitter motions at (source particle rest frame) time \(t=0\).  Carrying out the multiplication and applying the trigonometric identity \(\sin \alpha \sin \beta + \cos \alpha \cos \beta = \cos (\alpha - \beta)\) obtains

\begin{equation}
 \mbox{\boldmath$\beta$}_t \cdot   \mbox{\boldmath$\beta$}_{s,{\textnormal{{\scriptsize ret}}}} =  \cos \left[\gamma \omega( t - \mbox{\boldmath$\upsilon$} \cdot \mbox{\boldmath$r$}/c^2) - \omega t_r - \delta\phi_0\right]. 
\end{equation}

Substituting for \(t_r =  t - r_{\textnormal{{\scriptsize ret}}}/c \),

\begin{eqnarray}
 \mbox{\boldmath$\beta$}_t \cdot   \mbox{\boldmath$\beta$}_{s,{\textnormal{{\scriptsize ret}}}} = \nonumber \\ \cos \left[ (\gamma - 1) \omega t - \gamma \omega \mbox{\boldmath$\upsilon$} \cdot \mbox{\boldmath$r$}/c^2 + \omega (r_{\textnormal{{\scriptsize ret}}}/c) - \delta\phi_0 \right]. 
\label{retarded_modfact_gen}
\end{eqnarray}

Since the source zitter particle is stationary, \(r_{\textnormal{{\scriptsize ret}}}(t) = r(t) = r_0 - v_r t\), where \(r_0 \equiv r(t = 0)\) and \(v_r = c\beta_r\) is the component of the test zitter particle average velocity directly toward the source zitter particle,

\begin{eqnarray}
 \mbox{\boldmath$\beta$}_t \cdot   \mbox{\boldmath$\beta$}_{s,{\textnormal{{\scriptsize ret}}}} = \nonumber \\ \cos \left[ (\gamma - 1) \omega t - \gamma \omega \mbox{\boldmath$\upsilon$} \cdot \mbox{\boldmath$r$}/c^2 - \omega \beta_r t + \phi_r \right], 
\label{retarded_modfact_gen}
\end{eqnarray}

where  \(\phi_r \equiv \omega r_0/c - \delta\phi_0\) is a constant. 

For the case of circular motion of the test zitter particle around the source zitter particle, \(\mbox{\boldmath$\upsilon$} \cdot \mbox{\boldmath$r$}\) and \(\beta_r\) are zero and so

\begin{equation}
 \mbox{\boldmath$\beta$}_t \cdot   \mbox{\boldmath$\beta$}_{s,{\textnormal{{\scriptsize ret}}}} = \cos( (\gamma - 1) \omega t  + \phi_r ). 
\label{modfact_nonradial_vel} 
\end{equation}

Substituting for \(\omega = c/\lambdabar_0 = 2 c^2 m_e / \hbar \) obtains that  \(
(\gamma - 1) \omega = 2 c^2 m_e (\gamma - 1)/ \hbar \).  The frequency \(\nu_{\textnormal{{\scriptsize ret,nonrad}}}\) of the sinusoidal modulation of \(\mbox{\boldmath$\beta$}_t \cdot   \mbox{\boldmath$\beta$}_{s,{\textnormal{{\scriptsize ret}}}}\), for the retarded field acting on a nonradially-moving test zitter particle is then

\begin{equation}
\nu_{\textnormal{{\scriptsize ret,nonrad}}} = \frac{\omega(\gamma - 1)}{2\pi} = \frac{2 c^2 m_e (\gamma - 1)}{h}.
\end{equation}

The period \(T_{\textnormal{{\scriptsize ret,nonrad}}}\) of the sinusoidal modulation of \(\mbox{\boldmath$\beta$}_t \cdot   \mbox{\boldmath$\beta$}_{s,{\textnormal{{\scriptsize ret}}}}\) is thus

\begin{equation}
T_{\textnormal{{\scriptsize ret,nonrad}}} = \frac{1}{\nu_{\textnormal{{\scriptsize ret,nonrad}}}} = \frac{h}{2 c^2 m_e (\gamma - 1)},
\end{equation}

and the distance traveled by the test zitter particle in one period of the modulation is (using \(\gamma^{-2} = 1 - \beta^2\)),

\begin{eqnarray}
vT_{\textnormal{{\scriptsize ret,nonrad}}} = \frac{hv}{2 c^2 m_e (\gamma - 1)} \nonumber \\
= \frac{hv}{2 c^2 \gamma m_e (1 - \gamma^{-1})}  \nonumber \\
= \frac{hv(1 + \gamma^{-1})}{2 c^2 \gamma m_e (1 - \gamma^{-2})}  \nonumber \\= \frac{h(1 + \gamma^{-1})}{2 p},
\end{eqnarray}

where \( p \equiv \gamma m_e v\) is the test zitter particle  momentum. This differs from the de Broglie wavelength \(\lambda_{\text{de Broglie}} = h/p\) by the factor \((1 + \gamma^{-1})/2\), {\em i.e.},

\begin{equation}
vT_{\textnormal{{\scriptsize ret,nonrad}}} = \frac{1 + \gamma^{-1}}{2}\lambda_{\text{de Broglie}}.
\label{deBroglie_nonradial}
\end{equation}

Thus, in the low velocity limit where \((1 + \gamma^{-1})/2 \approx 1\), the de Broglie wavelength can be equated with the spatial period of the beat due to the difference of the source and test particles' zitter frequencies caused by time dilation. Therefore, the de Broglie wavelength in the low-velocity limit can be obtained as a modulation of the magnetic force caused by the zitterbewegung of one Dirac particle acting upon another of equal mass. At large velocity, \((1 + \gamma^{-1})/2 \approx 1/2\), and the modulation spatial period of the magnetic force acting on a (non-radially here) moving Dirac particle approaches half the de Broglie wavelength.

Therefore, in the low velocity limit, where \((1 + \gamma^{-1})/2 \approx 1\), the de Broglie wavelength can be equated with the spatial period of the beat due to the difference of the source and test particle zitter frequencies caused by time dilation.  Over the course of travel of one de Broglie wavelength, for non-radial low-velocity relative motion, the magnetic force due to the zitter motions corresponding to aligned spins will undergo one cycle of sinusoidal modulation.

For \(\beta_r\) nonzero, with the trigonometric identity \(\cos(\alpha + \beta) = \cos\alpha\cos\beta - \sin\alpha\sin\beta\),  Eq. (\ref{retarded_modfact_gen}) can be rewritten as

\begin{widetext}

\begin{equation}
 \mbox{\boldmath$\beta$}_t \cdot   \mbox{\boldmath$\beta$}_{s,{\textnormal{{\scriptsize ret}}}} = \cos((\gamma - 1) \omega t   - \gamma \omega \mbox{\boldmath$\upsilon$} \cdot \mbox{\boldmath$r$}/c^2) \cos(\omega \beta_r t - \phi_r) - \sin((\gamma - 1) \omega t  + \gamma \omega \mbox{\boldmath$\upsilon$} \cdot \mbox{\boldmath$r$}/c^2 ) \sin(\omega \beta_r t - \phi_r), 
\label{ret_two_freq} 
\end{equation}

\end{widetext}

illustrating that for non-zero radial motion it is not possible to factor \(\mbox{\boldmath$\beta$}_t \cdot   \mbox{\boldmath$\beta$}_{s,{\textnormal{{\scriptsize ret}}}}\) such that one factor is a sinusoidal modulation of frequency \((\gamma - 1) \omega\), as needed to obtain a modulation spatial period having the de Broglie wavelength in the low-velocity limit. For small relative velocity between the zitter particles and such that \(\beta_r >> \gamma - 1 \approx \beta^2/2\), radial motion leads to much more rapid oscillation of the inverse square law magnetic force between zitter particles than is the case for non-radial motion.  Therefore modulation of the inverse-square magnetic force due to radial relative motion is not obviously relatable to the de Broglie phase as is non-radial motion.  This is contradicted by observation, where, for example, the de Broglie phase is directly relatable to radial motion in the two-slit diffraction of electrons.  However up to this point only retarded electromagnetic interactions have been considered.

\section{Radial Relative Motion in the Time-Symmetric Electrodynamics Picture}
\label{sectexpdeborglie}

Although the sign reversals of the inverse-square-law magnetic force between zitter particles, that result due to small changes in interparticle separation compared to atomic scale, seem to rule out an exact correspondence with the Bohmian quantum force for \(s\) states, up to this point only time-retarded electromagnetic interactions have been considered. Since there is no physical basis for ruling out time-advanced interactions \emph{a priori}, it is worthwhile investigating what possible contribution they might make.  For this purpose the time-symmetric inverse-square magnetic interaction is considered for the case of constant-velocity relative motion between the test and source zitter particles. 

The modulating factor \(\mbox{\boldmath$\beta$}_t \cdot   \mbox{\boldmath$\beta$}_{s}\) for the case of a uniformly-translating test zitter particle moving in the advanced field of the source zitter particle is  

\begin{equation}
 \mbox{\boldmath$\beta$}_t \cdot   \mbox{\boldmath$\beta$}_{s,{\textnormal{{\scriptsize adv}}}} =  \cos \left[\gamma \omega( t - \mbox{\boldmath$\upsilon$} \cdot \mbox{\boldmath$r$}/c^2) - \omega t_a - \delta\phi_0\right],  
\end{equation}

where \(t_a\) is the time advancement from the test zitter particle at time \(t\) to the source zitter particle at the advanced time. Substituting for \(t_a =  t + r_{\textnormal{{\scriptsize adv}}}/c \),

\begin{eqnarray}
 \mbox{\boldmath$\beta$}_t \cdot   \mbox{\boldmath$\beta$}_{s,{\textnormal{{\scriptsize adv}}}} = \nonumber \\ \cos \left[ (\gamma - 1) \omega t - \gamma \omega \mbox{\boldmath$\upsilon$} \cdot \mbox{\boldmath$r$}/c^2 - \omega (r_{\textnormal{{\scriptsize adv}}}/c) - \delta\phi_0 \right]. 
\label{advanced_modfact_gen}
\end{eqnarray}

Since the source zitter particle is stationary, \(r_{\textnormal{{\scriptsize adv}}}(t) = r(t) = r_0 - v_r t\). The modulating factor similar to \(\mbox{\boldmath$\beta$}_t \cdot   \mbox{\boldmath$\beta$}_{s,{\textnormal{{\scriptsize ret}}}}\) of Eq. (\ref{retarded_modfact_gen}), for the case of a uniformly-translating test zitter particle moving in the advanced field of the source zitter particle is thus

\begin{eqnarray}
 \mbox{\boldmath$\beta$}_t \cdot   \mbox{\boldmath$\beta$}_{s,{\textnormal{{\scriptsize adv}}}} = \nonumber \\ \cos \left[ (\gamma - 1) \omega t - \gamma \omega \mbox{\boldmath$\upsilon$} \cdot \mbox{\boldmath$r$}/c^2 + \omega \beta_r t - \phi_a \right], 
\label{advanced_modfact_gen}
\end{eqnarray}

where  \(\phi_a \equiv \omega r_0/c + \delta\phi_0\) is a constant. Combining this result with Eq. (\ref{retarded_modfact_gen}) for motion in the retarded field, the modulating factor for the test zitter particle moving in the sum of the retarded and advanced fields is

\begin{widetext}

\begin{equation}
 \mbox{\boldmath$\beta$}_t \cdot   \mbox{\boldmath$\beta$}_{s,{\textnormal{{\scriptsize ret}}}} + \mbox{\boldmath$\beta$}_t \cdot   \mbox{\boldmath$\beta$}_{s,{\textnormal{{\scriptsize adv}}}} = \cos \left(\omega \left[ (\gamma - 1) t - \gamma  \mbox{\boldmath$\upsilon$} \cdot \mbox{\boldmath$r$}/c^2\right] + \phi_d\right)  \cdot \cos \left(\omega \beta_r t + \phi_s\right)
\label{TimeSymmetricFactor}
\end{equation}

\end{widetext}

where \(\phi_s = (\phi_r + \phi_a)/2 \) and \(\phi_d = (\phi_r - \phi_a)/2\) are constants.

Thus, summing the retarded and advanced magnetic radial inverse square zitter force components, the total can be written as a product of two sinusoidal factors.  This differs from the situation of the retarded-only force as illustrated by Eq. (\ref{ret_two_freq}), where such a factorization is not possible when there is radial relative motion.  Furthermore, for general uniform relative motion, the first sinusoidal factor of the right hand side of Eq. (\ref{TimeSymmetricFactor}) has an angular frequency of \(\omega (\gamma - 1)\), which as shown above leads to a modulation with a spatial period of the de Broglie wavelength in the limit of small relative velocity. Therefore, for general motion, unlike in the retarded-only case, the time-symmetric Coulomb-like magnetic force between zitter particles is sinusoidally modulated with spatial period equal to the de Broglie wavelength times \((1 + \gamma^{-1})/2\).

Also, in the time-symmetric picture, there is an additional electromagnetic force available that resembles the Coulomb force between stationary charges.  It is the electric force due to the advanced electric field.  Although time-symmetric electrodynamics usually assumes a time-symmetric field that is the mean of the retarded and advanced field, taking a difference of the advanced from the retarded field would agree well with Bohm's theory, at least in the \(s\) states.  Although this would cause the vanishing of the total time-symmetric Coulomb field of a stationary charge, such is not the case for the electric acceleration fields of relativistically circulating charges. In the case of a vanishing time-symmetric electrostatic force, the magnetic force considered here would need to play the role of Coulomb attraction.

\section{Discussion}
\label{sectdisc}

That the quantum force of Bohmian mechanics might be related to the magnetic force expected between charges undergoing relativistic circulatory motions on Compton wavelength scales seems consistent with recent work showing quantum potential energy can always be regarded as kinetic energy of concealed motion \cite{HollandY14}, and that the Bohmian quantum potential is an internal energy \cite{HileyY14}.  Also, the quantum potential has been identified directly with the zitterbewegung \cite{SalesiY9,SalesiRecamiY10}. If interacting charges are moving relativistically, then magnetic forces can be expected to be comparable in strength to Coulomb interactions and should not be overlooked.  

The restriction to equal-mass particles herein has limited the applicabilty of the present analysis to hydrogenic atoms consisting of equal-mass particles, such as the positronium atom. Consideration of the hydrogen atom itself will likely require regarding the proton properly as a composite of other spinning particles. The spin-spin coupling of the electron to the proton is very weak in the hydrogen atom because the proton intrinsic magnetic moment is three orders of magnitude weaker than the electron's, resulting in only the hyperfine splitting.  Nonetheless, if all charged particles must relativistically circulate on Compton wavelength scales, then  magnetic couplings that are stronger than usually accounted for can be expected.

Another problem of the present approach which must be overcome is that the radial inverse-square magnetic force tends to average to zero over a zitter period if the spins are opposite.  Going forward it will be attempted to obtain a deeper understanding by considering a more complete version of total inverse-square force, rather than just the explicitly radial term.  There are other inverse-square terms in Eq. (\ref{gen_zbw_mag_force}) which, although not explicitly radial, have radial contributions, and so need to be taken into account in analysis of the radial motion. Non-radial forces need to be investigated as well. 

Elementary particle internal periodicities and their modulations have recently been related to quantum behavior \cite{DolceY10,DolceY12,DolcePeraliY14}.  Electromagnetic forces deriving from acceleration fields of highly relativistic charges can plausibly mediate such periodicity modulations. Velocity fields may still play a role as well. For sub-luminal relativistic motion, electric and magnetic velocity fields remain finite but may be large in the plane of the zitterwegung motion.  For luminal motion, where the electromagnetic field becomes singular, a field-free approach based on a Fokker action principal obtains a well-defined dynamical picture \cite{DeLucaY10b}.

Since the early days of atomic physics, it has been recognized that the classical electrodynamics representation of the hydrogen atom as involving Kepler-like orbital motion with a correction for radiative decay is an approximation. The exact electromagnetic two-body problem was considered intractable \cite{Driver1969}, however, and propagation delay and magnetic effects due to the electron intrinsic spin and magnetic moment, when they became known, were considered to be negligible or small corrections. Nonetheless, it has been proposed that accounting properly for propagation delay, which requires reformulating the electromagnetic two-body problem in terms of functional rather than ordinary differential equations, might account for quantum behavior as a phenomenon emergent from classical electrodynamics \cite{RajuY5}. This mathematically rigorous approach leads to ``stiff'' equations of motion that can exhibit dynamical behavior on multiple scales, where zitterbewegung-like relativistic motions may be present simultaneously with non-relativistic average motions, and so may provide a classical electrodynamics explanation for the zitterwebegung motion that is a feature of the Dirac electron theory \cite{DeLucaY6,RajuRajuY8}

It has been suggested that physicality of time-advanced interactions might account for the apparent non-locality of quantum behavior \cite{Price1996,PriceWhartonY15a,PriceWhartonY15b}. The magnetic force found here is also present in the time-symmetric electrodynamics picture where the total field acting is the mean of the retarded and advanced fields \cite{Dirac1938,WheelerFeynman1945,Schild1963}.  In the case of a stationary charge, the time-symmetric field as the mean of retarded and advanced fields is identical to the retarded-only or undelayed Coulomb field. Similarly, the time-symmetric field of a stationary current loop is identical to the retarded-only field.  But, the time-symmetric field of the point charge circulating around a fixed point is non-constant, and distinct from the retarded-only field. The time-symmetric picture is thus physically distinct from the retarded-only case, and so it may be possible to determine whether time-advanced interactions are necessary to physical description of reality. Application of classical electrodynamics to explaining quantum behavior while taking time-symmetric account of delay may be found in \cite{DeLucaY9,DeLucaY10,DeLucaY12}. The de Broglie wavelength in the low-velocity limit is derived to order of magnitude in \cite{DeLucaY10}, without direct insertion of the Planck constant as in the present approach. In \cite{DeLucaY10}, furthermore, as in the present approach, non-instantaneous electromagnetic forces act time-symmetrically.

\section{Errata}
\label{errata} 

Previous versions of this article ({\em e.g.} \cite{LushY16}) accounted properly for time dilation \cite{Jackson_OC_TD} on the test particle zitter frequency as \(\omega' = \omega / \gamma\), where \(\omega\) is the zitter frequency in the test particle rest frame and \(\gamma = (1 - (v/c)^2)^{-1/2}\) is the Lorentz factor.  Although this is a correct description of the time dilation effect on frequency, it does not correctly account for the phase of the zitter motion of the non-stationary test particle as observed in the laboratory frame.  The latter is properly \cite{BaylisY7} determined by Lorentz transformation as  \(\omega \tau = \gamma \omega( t - \mbox{\boldmath$\upsilon$} \cdot \mbox{\boldmath$x$}/c^2)\), where \(\tau\) is the time coordinate in the test zitter particle rest frame, and \(\mbox{\boldmath$x$}\) is the  test zitter particle position. The present version (v13) of this article is corrected accordingly beginning at Eq. (\ref{first_correction}) above. The corrections do not negate the original conclusion of this article, that the time-symmetric electrodynamic interaction between Dirac particles leads to a modulation of a Coulomb-like magnetic force that has similarity to the de Broglie matter wave. The correction makes the stated relation closer, in that it naturally involves the relativistic momentum rather than its low-velocity approximation (as can be seen in Eq. (\ref{deBroglie_nonradial}) above, and by comparing with the corresponding equation, Eq. (26) of \cite{LushY16}.  The correction to zitter phase as described will have wider benefits when applied to the sequel \cite{LushY17} to the present article, which when corrected will be placed as indicated at \cite{LushY19}.

\section{Concluding Remarks}
\label{sectconc}

A preliminary evaluation of the magnetic interaction between  relativistically-circulating point charge particles has been performed.  It was found that the magnetic interaction between such particles has a radially-directed component with a strength equal to that of the Coulomb interaction, except that it is modulated by the relative orientation of particle circulatory motions, and their propagation-delayed phase difference. Taking account of relativistic time dilation due to motion of one of the particles, for non-radial relative motion, it was found that the spatial period of the modulation caused by the difference of zitterbewegung frequencies of the two particles can be equated with the de Broglie wavelength times \((1 + \gamma^{-1})/2\). For radial relative motion the de Broglie wavelength related modulation was found only if the electromagnetic field was time symmetric, {\em i.e.}, is the sum of retarded and advanced fields.  This type of interaction might find application in efforts to explain quantum behavior using only classical physics.

\appendix

\section{Evaluation of the Quantum Force in the Hydrogen Atom Ground State}
\label{appsectF_Q}

The quantum potential and resulting quantum force is calculated from the Schr\"odinger wavefunction. In the application of the Schr\"odinger equation to the hydrogen-like atom, the classical potential term is simply the Coulomb potential of a point charge, while the quantum potential depends on the solution. For the ground state solution of the time-independent Schr\"odinger equation, and for all {\em s}-states, the electron momentum vanishes \cite{HileyY14a}. In these cases, the quantum potential is everywhere opposite the Coulomb potential, and so the quantum force exactly cancels the Coulomb force.  However, the quantum force is not usually evaluated explicitly in the literature, since it is not needed to obtain the physical results of interest ({\em e.g., } that the electron momentum vanishes), so for completeness it is evaluated in the following for the hydrogen atom ground state. Also, it serves to illustrate that although the magnetic force component described here bears a similarity to the quantum force, it cannot be equated to it directly.  More generally, the quantum force cannot in principle be equatable to the magnetic force, because while the quantum potential is scalar in both its non-relativistic and relativistic \cite{HileyCallaghanY12} forms, the magnetic force must be derived from a vector potential.

The quantum potential is \cite{Bohm1952a_oc}

\begin{equation}
Q = -\frac{\hbar^2}{4m}\left[\frac{\nabla^2\rho}{\rho} - \frac{1}{2}\frac{(\nabla\rho)^2}{\rho^2} \right],
\end{equation}

where, if the Schr\"odinger wavefunction \(\psi \equiv R \exp(iS/\hbar)\), then \( \rho \equiv R^2\).

For the hydrogen ground state, the normalized Schr\"odinger wavefunction is \cite{GriffithsY8}

\begin{equation}
\psi = \left(\frac{2}{{a_0}^{3/2}} \right) e^{-r/a_0},
\end{equation}

or, equivalently,

\begin{equation}
\rho = \left(\frac{4}{{a_0}^{3}} \right) e^{-2r/a_0},
\end{equation}

and the quantum potential evaluates as

\begin{equation}
Q =  \frac{\hbar^2}{2m a_0} \left[\frac{2}{r} -  \frac{1}{{a_0}} \right],
\end{equation}

and the quantum force is

\begin{equation}
F_Q = -\nabla Q =  \frac{\hbar^2}{m a_0} \left[\frac{\hat{\mbox{\boldmath$r$}}}{r^2} \right].
\end{equation}

With \(a_0 = \hbar^2/m e^2\),

\begin{equation}
F_Q = e^2 \left[\frac{\hat{\mbox{\boldmath$r$}}}{r^2} \right].
\end{equation}

The quantum force thus exactly cancels the Coulomb attraction, in the Schr\"odinger hydrogen atom ground state.

%

%




\end{document}